\documentclass[useAMS,usenatbib]{mn2e}
\usepackage{epsfig}
\usepackage{subfigure}

\newif\ifAMStwofonts
\AMStwofontstrue

\def\be{\begin{equation}}
\def\ee{\end{equation}}

\def\mpc{\,{\rm {Mpc}}}

\def\kms{\,{\rm {km\, s^{-1}}}}

\def\cmd{\,{\rm {cm^{-2}}}}

\def\secu{\,{\rm {s^{-1}}}}
\def\eV{{\rm \ eV}}

\def\gtsima{$\; \buildrel > \over \sim \;$}
\def\ltsima{$\; \buildrel < \over \sim \;$}
\def\prosima{$\; \buildrel \propto \over \sim \;$}
\def\gsim{\lower.5ex\hbox{\gtsima}}
\def\lsim{\lower.5ex\hbox{\ltsima}}
\def\simgt{\lower.5ex\hbox{\gtsima}}
\def\simlt{\lower.5ex\hbox{\ltsima}}
\def\simpr{\lower.5ex\hbox{\prosima}}
\def\la{\lsim}
\def\ga{\gsim}

\def\Lya{Ly$\alpha$~}
\def\Lyb{Ly$\beta$~}

\newcommand{\CIV}{\mbox{C\,{\sc iv}}}
\newcommand{\CIII}{\mbox{C\,{\sc iii}}}

\newcommand{\NV}{\mbox{N\,{\sc v}}}

\newcommand{\OIII}{\mbox{O\,{\sc iii}}}
\newcommand{\OII}{\mbox{O\,{\sc ii}}}
\newcommand{\OI}{\mbox{O\,{\sc i}}}
\newcommand{\MgII}{\mbox{Mg\,{\sc ii}}}

\newcommand{\HI}{\mbox{H\,{\sc i}}}
\newcommand{\HeII}{\mbox{He\,{\sc ii}}}

\def\nh{{$n_{\rm H}$~}}

\def\LJ{{L$_{\rm J}$}}
\def\NHI{{$N($\HI$)$}}

\title[The QSO proximity effect with the FLO approach]{The QSO
  proximity effect at redshift $\langle z\rangle\simeq2.6$ with the FLO approach
  \thanks{Based on observations collected at 
  the  European  Southern  Observatory  Very  Large  Telescope,  Cerro
  Paranal,  Chile -- Programs  166.A-0106(A) and  during commissioning
  and science verification of UVES }}

\author[V. D'Odorico, et al.]{V. D'Odorico$^{1}$, M. Bruscoli$^{2,3}$, 
F. Saitta$^{1,2,4}$, F. Fontanot$^5$, M. Viel$^{1,6}$, 
\newauthor S. Cristiani$^{1,6}$ \&  P. Monaco$^2$ \\
$^1$INAF-OATS, Via Tiepolo 11, 34143 Trieste, Italy\\
$^2$Dipartimento di Astronomia, Universit\`a degli Studi 
di Trieste, Via Tiepolo 11, 34143 Trieste, Italy\\
$^3$INAF-IRA, L.go E. Fermi 5, I-50125 Firenze, Italy\\
$^4$European Southern Observatory, Karl-Schwarzschild-Str. 2, D-85748
Garching bei M\"unchen, Germany\\
$^5$Max-Planck-Institute for Astronomy, K\"onigstuhl 17, D-69117, 
Heidelberg, Germany\\
$^6$INFN/National Institute for Nuclear Physics, Via Valerio 2,
I-34127 Trieste, Italy\\} 
\pagerange{\pageref{firstpage}--\pageref{lastpage}}
\pubyear{2007}
\begin{document}

\maketitle
\label{firstpage}

\begin{abstract}
We revisit the proximity effect produced by QSOs at redshifts $2.1-3.3$
applying the FLO approach \citep{saitta} to a sample of $\sim 6300$
\Lya lines fitted in 21 high resolution, high signal-to-noise spectra.
This new technique allows to recover the hydrogen density field from
the \HI\ column densities of the lines in the \Lya forest, on the basis 
of simple assumptions on the physical state of the gas. 
To minimize the systematic uncertainties that could
affect the density recovering in the QSO vicinity, we carefully
determined the redshifts of the QSOs 
in our sample and modelled in detail their spectra to compute
the corresponding ionising fluxes. The mean density field obtained
from the observed spectra shows a significant over-density in the
region within 4 proper Mpc from the QSO position, confirming that QSOs
are hosted in high density peaks. The absolute value of
$\rho/\langle\rho\rangle$ for the peak is uncertain by a factor of
$\sim 3$, depending on the assumed QSO spectral slope and the minimum
\HI\ column density detectable in the spectra. 
We do not confirm the presence of a significant over-density extending
to separations of $\sim 15$ proper Mpc from the QSO, claimed in previous
works at redshifts $\langle z\rangle \simeq 2.5$ and 3.8.
Our best guess for the UV background ionisation rate based on the IGM mean 
density recovered by {\small FLO} is $\Gamma_{\rm UVB}\simeq 10^{-12}$ s$^{-1}$. 
However, values of $\Gamma_{\rm UVB} \simeq 3\times 10^{-12}$
s$^{-1}$ could be viable if an inverted temperature-density relation
with index $\alpha\simeq-0.5$ is adopted.
\end{abstract}
 
\begin{keywords}
intergalactic medium, quasars: absorption lines, cosmology:
observations, large-scale structure of Universe
\end{keywords}

\section{Introduction}
The ultraviolet radiation emitted by quasars (QSOs) is considered
the dominant source of  ionisation of the intergalactic medium (IGM)
at redshifts 2-4. 
Most of the absorption lines seen blue-ward of the \Lya emission in QSO
spectra (the so-called \Lya forest) are ascribed to fluctuations in
the low to intermediate density IGM \citep[see][for a recent
  review]{meiksin}.  
As a consequence, \Lya lines can be used as probes of the properties
and redshift evolution of the UV  ionising background.        

Observations show that the number density of \Lya lines increases with
redshift, but within single QSO spectra the number density of \Lya
lines decreases as the redshift approaches the QSO emission redshift.
This effect was first noticed by \citet{carswell82} and confirmed by
later studies \citep{murdoch,tytler}.  \citet{bajtlik} called this
deficiency of \Lya absorptions near the background QSO ``proximity
effect'' and attributed it to the increased ionisation of the \Lya clouds
near the QSO due to its ionising flux.  They used the proximity
effect in 19 low resolution QSO spectra to estimate the intensity of
the ultraviolet background radiation (UVB) at the Lyman
limit\footnote{The Lyman limit corresponds to the hydrogen ionisation
  energy $E_{\rm ion}=13.6\eV$ or $\lambda_{\rm LL}=912 $ \AA.},
$J_{\rm LL}$, for which they found the value $\log J_{\rm LL} \simeq
-21.0 \pm 0.5$ ergs cm$^{-2}$ sec$^{-1}$ Hz$^{-1}$ sr$^{-1}$ over the
redshift range $1.7<z<3.8$.  Several other authors carried out this
analysis on other data sets
\citep{lueal,kulkarni,bechtold,williger,cristiani95,giallo96}.
\citet{lueal} found the same result of \citet{bajtlik} in the same
redshift interval but using 38 QSO spectra.
\citet{bechtold}, with 34 low resolution QSO spectra covering the
range  $1.6<z<4.1$, found 
$\log J_{\rm LL}\simeq -20.5$. 
This value, 3 times larger than that of \citet{bajtlik} and
\citet{lueal}, was not confirmed by \citet{giallo96}.  
They obtained $\log J_{\rm LL}\simeq -21.3^{+0.08}_{-0.09}$, 
using 10 higher resolution ($R \sim 25000$) QSO spectra in the same redshift
interval as \citet{bechtold}.
More recently, \citet{scott00} considered a sample of 74
intermediate resolution QSO spectra in the redshift interval $1.7 < z
< 4.1$, from which they obtained a value $\log J_{\rm LL} \simeq
-21.1^{+0.15}_{-0.27}$.  
These authors paid a particular attention to the correct
estimate of the systemic redshift of the QSOs. Indeed, if a QSO
redshift lower than the true one is used in the analysis, the ionising effect
of the QSO on the \Lya clouds, and thus the derived value of $J_{\rm
  LL}$, are over-estimated.  

In the standard analysis of the proximity effect it is assumed that
the matter distribution is not altered by the presence of the QSO. The
only difference between the gas close to and far away from the QSO is
the increased  photoionisation rate due to the QSO emission.   
A consequence of this hypothesis is that there should be a correlation
between the strength of the proximity effect and the luminosity of the
QSO. However, observational results are not conclusive on this subject 
(e.g. \citealt{lueal,bechtold,srianand}; but see also
\citealt{liske01}). It is in fact likely that QSOs occupy over-dense
regions. Hierarchical models of structure formation predict that
super-massive black holes, that are thought to power QSOs, are in
massive halos \citep{granato,fontanot06,daangela} which are strongly
biased to high-density regions.  

The main aim of this work is to investigate the density distribution of
matter close to QSOs from the proximity effect. 
To this purpose, we applied the {\small FLO} (From Lines to
Over-densities) technique, developed in \citet[][Paper
  I]{saitta}, to a sample of 21 high resolution, high signal-to-noise
ratio QSO spectra.    
{\small FLO} converts the list of \HI\ column densities of \Lya lines
in a QSO spectrum into the underlying \nh hydrogen density field. This 
method significantly reduces the drawbacks of the line fitting
approach, in particular, the dependence on the fitting tool, the
subjectivity of the result, and the strong dependence of statistics on
the number of weak lines, which is in general poorly known. 
We could also constrain the value of the UVB ionisation rate, $\Gamma_{\rm
    UBV}$\footnote{$\Gamma_{\rm UVB} \equiv 4\pi \, \int_{\nu_{\rm
        LL}}^{\infty} \sigma_{\nu} (J_{\nu}/h\nu) d\nu$ where,
    $\sigma_{\nu}$ is the photo-ionisation cross section, $\nu_{\rm
      LL}$ is the frequency at the Lyman limit and $J_{\nu} = J_{\rm
      LL} (\nu/\nu_{\rm LL})^{-\gamma}$.}, by matching the recovered
  density field in the \Lya forest region with the mean cosmic density.


Similar analyses were performed in previous works using a
different approach, based on the determination of the cumulative probability
distribution function of pixel optical depth.  
\citet{rollinde} studied the density structure around QSOs using 
12 high resolution-spectra that belong also to our sample. These
authors marginally detected the presence of an over-density at
separations $3\la r \la 15$ proper Mpc, 
assuming an hydrogen ionisation rate $\Gamma_{\rm UVB} = 10^{-12}$
s$^{-1}$ (corresponding to $\log J_{\rm LL} \simeq -21.4$).
\citet{guimaraes}, using the same technique, investigated the
distribution of matter 
density close to 45 high-redshift ($z_{\rm em} \sim 3.8$) QSOs observed
at medium spectral resolution. Their study reveals gaseous
over-densities on scales as large as $\sim 15$ Mpc, with
higher over-densities for brighter QSOs.     

The paper is organised as follows: in Section~\ref{sectFLO}, we
introduce the {\small FLO} technique and describe an improvement to the
method. Section~\ref{sectOBS} reports the details of the observed data
sample, the QSO emission redshift and luminosity determination, and
the fitting analysis. 
In Section\ref{sect_PROX}, the proximity effect of QSOs on the
surrounding gaseous medium is pointed out and the over-density close
to QSOs is reconstructed with {\small FLO}. The characteristics of
the cosmological 
simulations used to obtain the mock spectra and the comparison with
observations are described in Section~6. Our conclusions
and the prospect for the future are listed in Section~7. 

Throughout this paper, we adopt a $\Lambda$CDM cosmology with the
following values for the cosmological parameters at $z=0$:
$\Omega_{\rm m}=0.26,\ \Omega_{\Lambda}=0.74,\ \Omega_{\rm b}=0.0463$,
$n_s=0.95$, $\sigma_8 = 0.85$ and $H_0 = 72$ km s$^{-1}$
Mpc$^{-1}$. These parameters are consistent with the best fits values
obtained from the latest results on the cosmic microwave background
\citep{WMAP} and from other flux statistics of the \Lya forest
\citep[e.g.][]{viel06}.


\section{The FLO technique}
\label{sectFLO}
{\small FLO} (From Lines to Over-densities) was introduced in Paper I.
The physical hypotheses at the base of this procedure are briefly
described in the following paragraphs.

Traditionally, the  analysis  of  the \Lya forest  was  based  on  the
identification and Voigt fit  of the  absorption lines in order to
derive the central redshift, the column  density and the Doppler
parameter (measuring the velocity dispersion in the line). 
This approach has two main drawbacks:  (i)  the  subjectivity of
the decomposition into components: the same complex absorption
can be resolved by different scientists (or software tools) in different
ways, both  in the number of components, and in the values of the
output parameters for a single component;  (ii)  the blanketing
effect of  weak lines: they can be  hidden by the stronger lines, so
that their exact number density is unknown and has  to be inferred
from statistical arguments. 
Unfortunately, since  the weak lines  are also the most  numerous, the
uncertainty  in their exact  number is  transformed into  a systematic
error of the computed statistical quantities. 

The {\small FLO} technique extends the line  fitting approach  by
identifying  a  new  statistical  estimator  describing the  physical
properties  of  the  underlying  IGM, the  hydrogen density
$n_{\rm H}$,  which is  linked to the  measured \HI\  column density
through the formula \citep{schaye}: 
\begin{eqnarray}
\label{tra}
N({\rm  HI})  \simeq & 3.7 \times 10^{13}\ \cmd
  (1+\delta)^{1.5-0.26\alpha} T_{0,4}^{-0.26}  \Gamma_{12}^{-1} \\
& \times \left(\frac{1+z}{4}\right)^{9/2} 
\left(\frac{\Omega_{\rm b}\,h^2}{0.024}\right)^{3/2}
\left(\frac{f_g}{0.178}\right)^{1/2},\nonumber
\end{eqnarray}
\noindent
where, $\delta  \equiv n_{\rm H}/\langle  n_{\rm H}\rangle -1$  is the
density contrast,  $T_{0,4} \equiv T_0  / 10^4$ K is the  temperature at
the  mean  density,  $\Gamma_{12}  \equiv \Gamma_{\rm UVB}/10^{-12}$
s$^{-1}$ is  the  H photo-ionisation rate due to the UV background,
$f_g  \approx \Omega_{\rm b}/\Omega_{\rm m}$ is 
the fraction of the mass in gas and $\alpha$ is the index of the
temperature-density relation for the IGM which depends on the ionisation
history  of the  Universe.   Equation~\ref{tra} relies  on three  main
hypotheses:  (i)  \Lya  absorbers   are  close  to  local  hydrostatic
equilibrium, i.e.  their characteristic  size will be typically of the
order  of  the   local  Jeans  length  (\LJ), which can be
approximated as\footnote{Note that the formula in the original papers
  has the wrong sign for the exponent of $\alpha+1$.} \citep[][]{nusser,zaroubi}:  
\begin{eqnarray}
\label{LJ}
L_{\rm J} \simeq \\
& 1.498\left(\frac{\Omega_{\rm m}h^{2}}{0.135}\right)^{-1/2}\left(\frac{T_{0,4}}
{1.8}\right)^{1/2}  
 \left(\frac{\alpha + 1}{1.6}\right)^{1/2}\left(\frac{1+z}{3.5}\right)^{-1/2}
\nonumber 
\end{eqnarray}
in comoving Mpc, where  $h \equiv H_{0}/100$ km s$^{-1}$ Mpc$^{-1}$
and  the other parameters were already defined;  (ii)   the  gas
is  in photo-ionisation equilibrium;   (iii)   the   `effective equation   of
state' \citep{huignedin}, 

\be
T=T_0\,(\delta +  1)^{\alpha}, 
\label{eqstate}
\ee
holds for  the optically-thin IGM gas. 

In order to  apply eq.~\ref{tra} we have, first of  all, to go through
the  Voigt fitting process  of the  \Lya forest  absorptions in  a QSO
spectrum.  Then,  to transform  the list of  \HI\ column  densities of 
\Lya lines into the matter density field which generated them, we have
to perform the following steps: 
\par\noindent (1) group adjacent \Lya lines into absorbers of size  of 1 \LJ\
with column density equal to  the sum of column densities and redshift
equal to the weighted average  of redshifts, using column densities as 
weights.     
The absorbers  are created with a friend-of-friend algorithm:
\begin{enumerate} 
\item the  spatial separation between  all the possible line  pairs is
computed and the minimum separation  is compared with \LJ, computed at
the \NHI-weighted redshift mean of the pair;
\item if  the two lines  of the pair  are more distant than  the local
\LJ,  they  are classified  as  two  different  absorbers, stored  and
deleted from the line list;
\item  if the  two  lines are  closer  than the  local  \LJ, they  are
replaced in  the line list  by one line  with a redshift equal  to the
\NHI-weighted mean of the two  redshifts and a column density equal to
the sum of the two column densities;
\item the procedure is iterated until all the lines are converted into
absorbers.
\end{enumerate}
\par\noindent (2) transform the  list of column densities of absorbers
into a list of $\delta$ inverting  eq.~\ref{tra};
\par\noindent (3)  bin the redshift  range covered by the  \Lya forest
into  steps of 1  \LJ\ and  distribute the  absorbers onto  this grid,
proportionally to  the superposition  between absorber size  (which is
again 1  \LJ) and bin.  Two cases are considered for the treatment of
the empty bins: in the `lower limit' case the bin is filled with an
absorber of null column density (corresponding to $\delta=-1$), while
in the `upper limit' case the bin is  filled with an  absorber with
hydrogen  density  contrast corresponding  to  the minimum  detectable
column density in our data, $\log N($\HI$) = 12\ \cmd$, at the redshift
of the bin. 

For the IGM parameters in eqs.~\ref{tra} and \ref{LJ} we adopted the
same values of paper I:  $T_{0,4}=1.8$, $\Gamma_{12}=1$ and
$\alpha=0.6$. 
  
     
\begin{figure}
\begin{center}
\includegraphics[width=8cm]{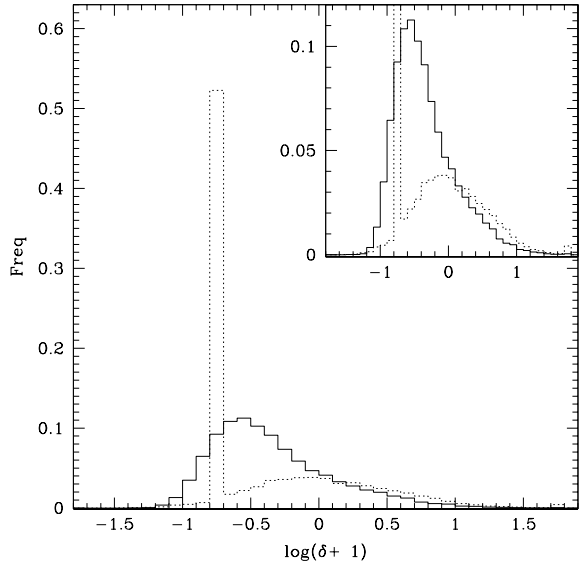}
\caption{{\protect\footnotesize{Distribution of the fraction of bins 
as a function of their $\log(\delta+1)$-value for the true field (solid line)
and the recovered one (dotted line) from the simulated lines of
sight in the upper limit case. The smoothing scale is the Jeans length.}}}  
\label{distrib_lj}
\end{center}
\end{figure}
     
\begin{figure}
\begin{center}
\includegraphics[width=9cm]{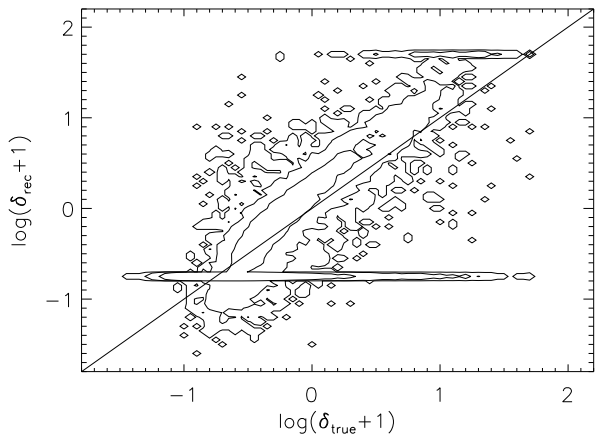}
\caption{{\protect\footnotesize{Contour scatter plot of the true
      versus reconstructed $\delta$ fields from simulations in the
      upper limit case. The
      smoothing scale is the Jeans length. The
      contours show the number density of pixels which increases by a
      factor of 10 at each level. }}}  
\label{contour_lj}
\end{center}
\end{figure}

\subsection{An improvement of the {\small FLO} method}

In paper I, we tested the quality of the density reconstruction
of {\small FLO} using synthetic QSO spectra drawn from a cosmological
simulation (the same that is described in Section~6). 
Figure~\ref{distrib_lj} and \ref{contour_lj} shows the results of those
tests. Both the true density field and the reconstructed one 
along the simulated lines of sight were rebinned into steps of 1 Jeans
length . The recovered density field from the simulated lines of
sight presents two main problems: (i) most of the points of the true
density field belonging to under-dense regions are not recovered with
the correct $\delta$-value, they are accounted for as empty bins or,
for those approaching the average density,  
their value is over-estimated and they are moved to moderately
over-dense regions; (ii) the number of points in the over-dense
regions is over-estimated due to a systematic assignment of
larger-than-true $\delta$-values to those points.  
The latter effect causes also an over-estimate of the average
$\delta$ of the distribution which is $40-50$ percent higher
than that of the true density field.  
In paper I, we solved this problem by normalising both the true and the
reconstructed $\delta$-field in order to have the same mean value,
$\langle \delta + 1\rangle  = 1.0$. This solution is not viable for
the present analysis, so we looked for the physical reasons of the
wrong reconstruction in order to improve the performances of {\small
  FLO}. 

The failure in reproducing the under-dense regions is likely due to
the fact that gas below the average density is still expanding and our
primary hypothesis, the local hydrostatic equilibrium, cannot be
applied. There is no simple solution to this problem, however we have
already tested in paper I that under-dense regions have a negligible
effect on statistical quantities. 

The problem regarding the moderately over-dense regions depends on the
scale adopted for the reconstruction of the absorbers, the Jeans
length \LJ, which appears to be too large. As a consequence, absorbers
have on average too large column densities 
which translate into an over-estimate of the $\delta$-values for the
over-dense regions. 
\begin{figure}
\begin{center}
\includegraphics[width=8cm]{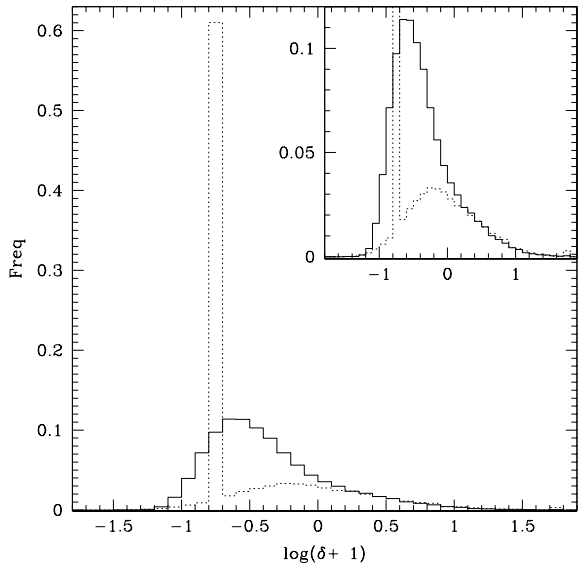}
\caption{{\protect\footnotesize{As in Fig.~\ref{distrib_lj} but using
      a filtering scale L$_{\rm F} \sim 0.7$~\LJ.}}}  
\label{distrib_lf}
\end{center}
\end{figure}
\begin{figure}
\begin{center}
\includegraphics[width=9cm]{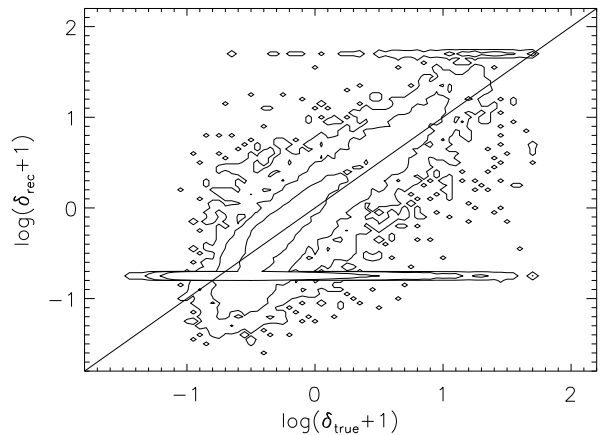}
\caption{{\protect\footnotesize{As in Fig.~\ref{contour_lj} but using
      a filtering scale L$_{\rm F} \sim 0.7$~\LJ.}}}  
\label{contour_lf}
\end{center}
\end{figure}
The need for a smoothing length smaller than \LJ\ to better reproduce the
density field traced by the \Lya forest, was discussed in detail by 
\cite{gnedin_hui}. They affirmed that the correct filtering scale
depends on the re-ionisation history of the Universe and is in general
smaller than the Jeans scale after re-ionisation, and larger prior to
it. We used eq.~A4 of \cite{gnedin_hui} to compute  L$_{\rm F}$, adopting
$z_{\rm rei} = 11$ and $z_{\rm obs}  = 2.5$. The results using the
filtering scale L$_{\rm F} \simeq 0.7$~\LJ\ are shown in
Fig.~\ref{distrib_lf} and \ref{contour_lf}.  
The moderately over-dense regions are now
successfully recovered both in the number of points and in the
$\delta$-values, and the average $\delta$ of the distribution is in
agreement with the true one within $10$ percent.   
We verified that varying the re-ionisation redshift or the
  temperature at the mean density by $\sim 20$ percent does not have 
a significant effect on the density reconstruction process.

In the following analysis, the filtering scale L$_{\rm F}$ replaced
the Jeans length in the {\small FLO} algorithm described before. 


\section{Observed data sample}
\label{sectOBS}

\begin{table*}
\begin{center}
\begin{minipage}{170mm}
\caption{Relevant properties of the QSOs forming our sample. See text for
  further details. }
\label{tab1}
\begin{tabular}{@{} l l l c c c c c c c c c}
\hline  
QSO& $z_{\rm em}$(Ref.) & line & $\Delta z_{\rm Ly\alpha}$ & $b_J$ &
$r_F$ &  \multicolumn{3}{c}{$\Gamma / 10^{40}$} & $\log L_{\rm
  LL}$ & $r_{\rm eq}$  \\
& & & & & &  F07 & T02 & HM && (Mpc) \\     
\hline
HE1341-1020$^a$ & 2.142(1) & \MgII & 1.6599-2.142 & 18.68 & 17.52 &
0.11$\pm$0.03 & 0.10$\pm$0.02 & 0.12$\pm$0.02 & 30.712 & 2.9 \\
Q0122-380    & 2.2004(2)  & H$\beta$ & 1.709-2.2004 & 17.34 & 16.70 &
0.8$\pm$0.3 & 0.5$\pm$0.1 & 0.6$\pm$0.1 & 31.267 & 5.5 \\
PKS1448-232  & 2.224(1)  & \MgII & 1.729-2.224 & 17.09 & 16.87 & 1.0$\pm$0.3
& 0.56$\pm$0.09 & 0.6$\pm$0.1 & 31.377 & 6.3 \\
PKS0237-23   & 2.233(1) & \MgII & 1.737-2.233 & 16.61 & 16.21 & 1.3$\pm$0.6
& 0.8$\pm$0.2 & 1.0$\pm$0.2  & 31.573 & 7.9 \\
J2233-606    & 2.248(1)  & \OI & 1.7496-2.248 & 16.97 & 17.01 & 1.4$\pm$0.2
& 0.66$\pm$0.09 & 0.8$\pm$0.1  & 31.437 & 6.7 \\
HE0001-2340  & 2.265(1) & \MgII & 1.764-2.265 & 16.74 & 16.46 & 1.2$\pm$0.5
& 0.7$\pm$0.2 & 0.9$\pm$0.2  & 31.538 & 7.6 \\
HE1122-1648$^b$  & 2.40(3)  & & 1.878-2.344 & 16.61 & 16.32 & 1.8$\pm$0.5 &
1.0$\pm$0.2 & 1.2$\pm$0.2 & 31.679 & 8.9 \\
Q0109-3518   & 2.4057(4)  & [\OIII] & 1.883-2.4057 & 16.72 & 16.37 &
1.8$\pm$0.5 & 1.0$\pm$0.2 & 1.1$\pm$0.2 & 31.640 & 8.5 \\
HE2217-2818$^b$  & 2.412(1) & & 1.888-2.355 &16.47 & 16.16 & 2.7$\pm$0.5 &
1.3$\pm$0.2 & 1.5$\pm$0.2 & 31.744 & 9.6 \\
Q0329-385    & 2.437(5) & \MgII & 1.9096-2.437  & 17.20 & 16.91 &
1.6$\pm$0.3 & 0.7$\pm$0.1 & 0.8$\pm$0.1 & 31.472 & 7.0 \\
HE1158-1843$^{a,b}$ & 2.448(1) & & 1.919-2.391 &17.09 & 16.84 & 1.3$\pm$0.2
& 0.70$\pm$0.09 & 0.8$\pm$0.1 & 31.525 & 7.4 \\
HE1347-2457$^a$ & 2.5986(4) & H$\beta$ & 2.046-2.5986 & 17.35 & 16.14
& 0.7$\pm$0.3 & 0.7$\pm$0.1 & 0.8$\pm$0.1 & 31.542 & 7.6 \\
Q0453-423       & 2.669(1)  & \OI & 2.106-2.669 & 17.69 & 16.74 &
0.7$\pm$0.4 & 0.5$\pm$0.1 & 0.6$\pm$0.1  & 31.451 & 6.8 \\
PKS0329-255$^b$  & 2.696(1) & & 2.129-2.635 &17.88 & 17.71 & 0.8$\pm$0.1 &
0.38$\pm$0.06 & 0.44$\pm$0.07 & 31.396 & 6.4 \\
HE0151-4326$^b$  & 2.763(1)  & & 2.186-2.701 &17.48 & 16.93 & 1.7$\pm$0.4 &
0.9$\pm$0.1 & 1.0$\pm$0.1 & 31.594 & 8.1 \\
Q0002-422    & 2.769(1) & \OI & 2.191-2.769 & 17.50 & 16.89 & 1.5$\pm$0.4 &
0.8$\pm$0.1 & 1.0$\pm$0.1  & 31.589 & 8.0 \\
HE2347-4342$^a$ & 2.880(1) & \OI & 2.285-2.880 & 17.12 & 16.30 & 2.0$\pm$0.7
& 1.3$\pm$0.2 & 1.5$\pm$0.2 & 31.810 & 10.4 \\
HS1946+7658$^a$     & 3.058(6) & \OI & 2.435-3.058 & 16.64 & 15.76 & 5$\pm$1
& 2.7$\pm$0.4 & 3.1$\pm$0.5 & 32.118 & 14.8 \\
HE0940-1050 & 3.0932(4) & H$\beta$ & 2.465-3.0932 & 17.08 & 16.08 &
3$\pm$1 & 2.0$\pm$0.3 & 2.3$\pm$0.3 & 31.972 & 12.5 \\
Q0420-388$^a$ & 3.1257(1) & \OI & 2.493-3.1257 & 17.44 & 16.70 & 2.6$\pm$0.5
& 1.2$\pm$0.2 & 1.4$\pm$0.2 & 31.848 & 10.8 \\
PKS2126-158 & 3.292(7) & [\OIII] & 2.633-3.292 & 17.54 & 16.37 & 4$\pm$1
& 1.8$\pm$0.3 & 2.0$\pm$0.3 & 31.958 & 12.3 \\
\hline
\end{tabular}

\footnotesize{$^a$ QSOs with associated absorption systems; $^b$ QSO
  not considered in the proximity study. \\ 
References: (1) this paper; (2) \citealt{sulentic}; (3)
  \citealt{kim02}; (4) P. Marziani, private comm.; (5)
  \citealt{espey89};  (6) \citealt{fan_tytler}; (7) \citealt{scott00}.
}
\end{minipage}
\end{center}
\end{table*}

\begin{figure}
\begin{center}
\includegraphics[width=8cm]{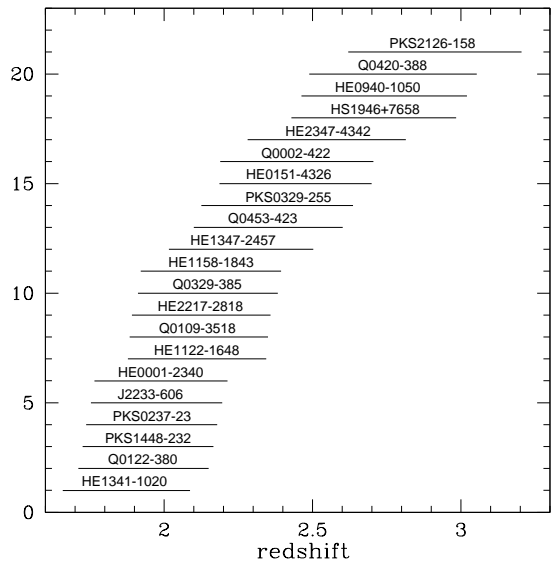}
\caption{\Lya forest redshift coverage  of the QSOs in our 
  sample. }
\label{qsodist}
\end{center}
\end{figure}

Most of  the observational data used  in this work were obtained
with the UVES spectrograph \citep{dekker} at the Kueyen unit of the
ESO VLT (Cerro Paranal,  Chile) in  the framework of  the ESO Large
Programme (LP): ``The Cosmic Evolution of the IGM''
\citep{bergeron04}.  Spectra of 18 QSOs were obtained in service
mode with the aim of studying the physics of the IGM in the redshift
range 1.7-3.5. The spectra have a resolution $R \sim 45000$ and a
typical signal to noise ratio (SNR) of $\sim  35$  and 70  per  pixel
at  3500  and  6000 \AA,  respectively. 
The wavelength range goes from 3000 to 10,000 \AA, except for two  intervals  
of about 100 \AA, centred at $\sim 5800$ and $8600$ \AA\  where the signal is
absent, due to the gap between the two CCDs forming the red mosaic. 
In the spectra of the two QSOs at higher redshift (Q0420-388 and
PKS216-158) there are three gaps centred at $\sim 5640$ and $8600$
\AA\ of about 100 \AA\ and at $\sim 6680$ \AA\ of width $\sim 50$ \AA.   
Details of the data reduction can be found in \citet{chand04} and
\citet{aracil04}. In particular, the continuum level was estimated
with an automatic iterative procedure 
which underestimates the true
continuum in the \Lya forest to about 2 percent at $z \sim2.3$. In the
process of fitting the lines in the \Lya forest, we corrected the
continuum level in the spectral intervals where it was clearly underestimated by
interpolating the regions free from absorption with polynomials of 3rd order. 

We  added to the main sample 3 more QSO spectra with comparable
resolution and SNR: 
\par\noindent  -  J2233-606 \citep{j2233}.  Data  for  this QSO  were
acquired during the commissioning of UVES in October 1999.
\par\noindent -  HE1122-1648 \citep{kim02}.   Data for this  QSO were
acquired during the science verification of UVES in February
2000. The reduced and  fitted spectrum was kindly provided  to us by Tae-Sun
Kim.
%
\par\noindent -  HS1946+7658 \citep{kirk97}.  Data for this  QSO were
acquired with Keck/HIRES in July 1994. 
%

In the following sub-sections, the procedure to derive the QSO
emission redshift and luminosity and the \Lya line lists is
described. 
Table~\ref{tab1} reports for each QSO in the sample the final emission
redshift and how it was computed, the studied \Lya redshift
range, the apparent magnitude of the QSO, and the corresponding H
ionisation rate resulting from the three adopted QSO spectra, the
luminosity at the Lyman limit and the radius of influence of the QSO
ionising flux.  
Figure~\ref{qsodist} shows  the distribution  in redshift of  the \Lya
forests for  all the QSOs of  the sample. We considered the range
between 1000 km s$^{-1}$ red-ward the \Lyb emission, to avoid
contamination by \Lyb absorption lines, and the \Lya emission. 

\subsection{Estimate of the QSO systemic redshifts}
\label{sub_zem}

The knowledge of the correct systemic redshift of the QSO is of
fundamental importance when using the proximity effect to estimate 
both the intensity of the UV ionising background and the density
structure close to the QSO itself. 

Emission redshift of QSOs at $z_{\rm em}\ga1.5$ are generally computed
by the positions of the most prominent UV emission lines, in
particular \HI\ \Lya $\lambda\,1216$ and \CIV\ $\lambda\,1549$.
However, it was assessed by several studies
\citep[e.g.][]{gaskell,wilkes86,espey89,corbin,tytler_fan,boroson92,laor95,marziani96,mcintosh99,vandenberk01,sulentic}
that high-ionisation emission lines (e.g., \CIV, \NV\ $\lambda\,1240$,
\CIII] $\lambda\,1909$, and \HI\ \Lya) are on average blue-shifted by
  several hundreds of km s$^{-1}$ with respect to low-ionisation lines
  (e.g., \OI\ $\lambda1304$, \MgII\ $\lambda2798$) and the permitted
  \HI\ Balmer series. In contrast, redshifts from narrow forbidden
  lines (e.g., [\OII] $\lambda\,3727$, [\OIII] $\lambda\,5007$) are
  observed to be within 100 km s$^{-1}$ of the broad \MgII\ and Balmer
  lines. 
Furthermore, in local active galactic nuclei, redshifts from narrow
forbidden lines showed agreement to $\sim100$ km s$^{-1}$ of the
accepted systemic frame determined by stellar absorption features and
\HI\ 21 cm emission in the host galaxies \citep{gaskell,vrtilek,hutchings}.    
  
In order to estimate the correct redshift, we carried out a detailed analysis of the
emission lines of the QSOs in our sample both using data from
the literature and directly fitting the lines in the UVES spectra. 
In decreasing order of precision, we adopted the redshifts measured
by: i) narrow forbidden lines, mainly [\OIII],
ii) H$\beta$, iii) \MgII, and iv) \OI. There are 5 QSOs in the sample
for which none of these lines was measured. We excluded those objects
from the proximity study, disregarding the portion of the spectrum within
5000 km s$^{-1}$ of the best estimate of the emission redshift
(generally obtained from \Lya\ and/or \CIV\ emission lines).  

The determination of the redshifts from the UVES spectra were obtained by
re-binning the region of the emission, normalising it to the local
continuum and fitting a Gaussian profile to the line.  
Table~\ref{tab1} gives the QSO redshifts and the details of the
estimate. 


\subsection{Estimate of the QSO ionising fluxes} 
\label{sub_lum}

Getting closer and closer to the QSO, the UV ionising field becomes
dominated by the intrinsic QSO emission flux. In order to derive the
matter density distribution around the QSO using the
observed variation of the absorption features in the QSO spectrum, a
reliable determination of the intrinsic luminosity of the QSO has to
be obtained. 

Magnitudes of the objects in our sample were taken from the
GSC-II catalogue \citep{GSC2} and are reported in Tab.~\ref{tab1}.  
In order to estimate the corresponding intrinsic bolometric
luminosity, we adopted the following procedure.  

We considered the QSO template library defined in \citet{fontanot},
based on high quality SDSS QSOs spectra in the redshift 
interval $2.2 < z < 2.25$. This redshift range was chosen in 
order to maximise the level of completeness of the sample and the
wavelength interval long-wards of the \Lya emission. Moreover, in this
redshift range the dynamical response of the SDSS spectrograph is
such that the \Lya line is completely sampled in all spectra. 
In the original paper, the authors considered the rest-frame spectra
of the 215 QSOs forming the final sample and they used a continuum
fitting technique in order to extend the information blue-ward of the
\Lya.  A mean continuum slope $\gamma = 0.7 \pm 0.3$ was obtained for
the objects in the library, where $f_\nu \propto \nu^{-\gamma}$ and
$f_\nu$ (corresponding to $4\pi\,J_{\nu}$) is the QSO flux in units of
ergs cm$^{-2}$ $\secu$ Hz$^{-1}$. \citet{fontanot} demonstrated that
this library is suitable 
for predicting QSO colours up to $z \simeq 5.2$. 

We used the template spectra in the library to compute a
synthetic $b_J$ and $r_F$ magnitude at each emission redshift
listed in Table~\ref{tab1}.  For reproducing the $b_J$ and $r_F$
photographic magnitudes, the response of the spectral code {\small
  PEGASE} \citep{pegase} was assumed.  Then, the templates were 
renormalised in each band separately, by requiring the synthetic
magnitude to match the observed one, and the renormalised spectra were 
used to give a prediction for the AB magnitudes at 912 \AA\ 
($M_{912}^{b_J}$ and $M_{912}^{r_F}$ respectively). 
The quantity $\Delta M_{912} = M_{912}^{b_J}-M_{912}^{r_F}$ was
adopted as an estimator of the agreement between the slope of the
template and the intrinsic slope of the considered QSO: we then associated 
to each observed QSO the template
with the smaller $\Delta M_{912}$. For 18 out of 21 QSOs in our
sample this procedure gave $\Delta M_{912}$ values lower than $0.001\ 
mag$. For the remaining 3 objects the values were respectively $0.1$
(PKS0329-255), $0.2$ (HE1347-2457) and $0.4\ mag$ (HE1341-1020). In
these cases, there was no template in the library which reproduced the
correct intrinsic slope of the observed QSO.  
The $\Delta M_{912}$ can be taken as a measure of the systematic error 
on the luminosity computed for these 3 objects.

The selected and renormalised template spectra were completed in the
region blue-ward of the \Lya emission using three possible
extrapolations:  
\begin{description}
\item[F07:] the continuum slope of the template spectrum red-ward
  of the \Lya emission;
\item[T02:] a fixed power law with slope $\gamma=1.8$
  \citep{madau99,telfer}; 
\item[ HM:] a fixed power law with slope $\gamma =1.5$
  \citep{HM}. 
\end{description}
Then, the spectra were used to estimate the QSO monochromatic
luminosities, $L_\nu$ (ergs $\secu$ Hz$^{-1}$). 
The hydrogen photo-ionisation rates due to the QSO radiation were
obtained through the formula:  

\be
\Gamma_{\rm QSO} [\rm{s}^{-1}]=\frac{1}{4\pi r_{\rm abs}^2} \Gamma
\label{eqGQ} 
\ee
where $r_{\rm abs}$ is the distance (in cm) between the QSO emission
redshift and the absorber and $\Gamma$ is obtained by integrating the
monochromatic luminosity in the wavelength range $300 <
\lambda < 912$ \AA: 

\begin{equation}
\Gamma =  \int_{\nu_{LL}} \, \frac{L_{\rm \nu}}{h \nu} \, \sigma_\nu \, d \nu
\end{equation}
where $\sigma_\nu$ is the absorbing cross-section for neutral hydrogen
\citep{osterbrock}. 

The error on $\Gamma$ was computed by taking into account the
uncertainties in the QSO magnitudes.
We randomly extracted $100$ values of $b_J$ and $r_F$ in the corresponding
allowed range and we repeated the previous procedure. The obtained
mean $\Gamma$ values are in good agreement with the estimate based on
the observed magnitudes, and we adopted the variance on the 100
realizations as the error on $\Gamma$. 
The values of $\Gamma$ with their uncertainties, for the three
investigated QSO continua, are reported in Table~\ref{tab1}.
The three estimates F07, T02 and HM agree within $2\,\sigma$. In particular, the $\Gamma$ for
the T02 and HM models are always within $1\,\sigma$. As
a consequence the differences between the results obtained adopting
the three $\Gamma$ values could be interpreted also as due to the
uncertainties in the $\Gamma$ itself.  
    
The radius of the sphere of influence of each QSO in our sample is
obtained by relating the intensity of the UV ionising
background at the Lyman limit, $J_{\rm LL}$, and the luminosity of the
QSO at the same frequency, $L_{\rm LL}$, measured as described above,   

\be
r_{\rm eq}=\frac{1}{4\pi} (L_{\rm LL}/J_{\rm LL})^{1/2}
\ee
The luminosities depend slightly on the adopted slope, however the
differences are small so we used the average of the three values to
compute $r_{\rm eq}$. 
For the UV background, we adopted the value $J_{\rm LL}\sim 4\times
10^{-22}$ ergs $\cmd$ $\secu$ Hz$^{-1}$ sr$^{-1}$, corresponding to an
ionisation rate $\Gamma_{\rm UVB} \simeq 10^{-12}$ $\secu$. The
resulting average $L_{\rm LL}$ and the radii expressed in proper Mpc
are reported in Table~\ref{tab1}. 

\subsection{Compilation of the line lists}
\label{sub_fit}

All the lines in the \Lya regions of the LP QSOs plus J2233-606 were
fitted with the {\small FITLYMAN} tool \citep{font:ball} of  
the {\small  ESO  MIDAS}  data reduction
package\footnote{http://www.eso.org/midas}. In the case of 
complex saturated  lines, we used the minimum number of  components to
reach $\chi^{2} \le  1.5$.  Whenever possible, the other  lines in the
Lyman series were used to  constrain the fit.  The minimum \HI\ column
density detectable at  3\,$\sigma$ with the SNR of  the spectra of our
sample is $\log N($\HI$)\simeq 12$ cm$^{-2}$.

Metals in  the forest were  identified and the  corresponding spectral
regions were masked to avoid effects of line blanketing. The same
treatment was given to \Lya lines with column density
$N($\HI$)\ge 10^{16}$ cm$^{-2}$ since, on the one hand, strong \HI\
lines can hide weaker lines as much as metal lines, and on the other
hand the application of the {\small FLO} algorithm is valid only in the linear
or mildly non-linear regime, or for over-densities $\delta \sim {\rm
  few}\times 10$.   
We eliminated \Lya lines with Doppler  parameters $b \le 10$ km s$^{-1}$,
that are  likely unidentified metal absorptions.  They represent about
the 7 percent  of the total sample of \Lya lines.   The output of this
analysis is a  list of \Lya lines for each  QSO with central redshift,
\HI\ column  density,  Doppler   parameter  and  the  corresponding  errors
obtained with {\small FITLYMAN}.

The \Lya forests of the remaining two QSOs, HE1122-1648 and
HS1946+7658, were fitted with the {\small
  VPFIT}\footnote{http://www.ast.cam.ac.uk/$\sim$rfc/vpfit.html} 
package. The same constraints on Doppler parameter and column density
were applied to obtain the final \Lya\ line lists for these
QSOs. The difference in the {\small FLO} results due to the different
fitting software, {\small FITLYMAN} and {\small VPFIT}, is negligible
and was discussed in Paper I.

\section{The proximity effect}
\label{sect_PROX}

\begin{figure}
\begin{center}
\includegraphics[width=8cm]{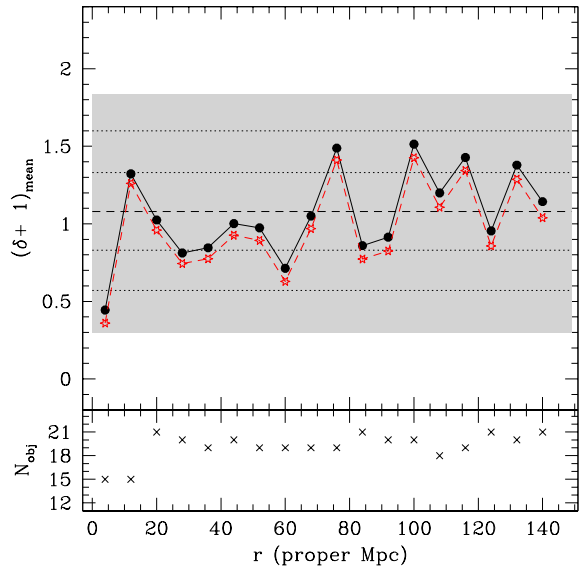}
\caption{{\protect\footnotesize{The mean of the $\delta$ field reconstructed 
around the QSOs in our sample,  neglecting the effect of the UV flux from the QSO themselves. 
The solid dots represent the upper limit case while the open stars trace
      the lower limit case (see text). The shaded area marks the
      $3\,\sigma$ region for the upper limit case, the dotted lines
      mark the corresponding 1 and $2\,\sigma$ levels and the dashed
      line the average IGM density contrast. The lower panel shows the
      number of objects 
      contributing to each bin.}}}  
\label{fig08}
\end{center}
\end{figure}

Moving closer to the QSO a decrease of the number density of \Lya
absorption lines with respect to the mean value is expected. 

We applied {\small FLO} to recover the density field in the
neighbourhood of the QSOs in our sample\footnote{Note that due to
  the uncertainty in the emission redshifts, 5 QSOs were excluded from
  the computation of the proximity effect.}, deliberately neglecting
the contribution of the QSO radiation field to the UV ionising flux.  
The density field traced by the \Lya forest was computed in each QSO
spectrum of our sample. The data were 
re-binned into bins of 8 proper Mpc in length. Bins covered by more
than 33 percent by masked intervals were eliminated from the final
count for the single object.  Then, in each bin, the mean of all the
$\delta$ values contributing to that bin (one per QSO at maximum) was
computed. 
The resulting $\delta$-field is shown in Fig.~\ref{fig08} together
with the number of QSOs contributing to each bin. The upper and
lower limit case are shown, the difference between the two
reconstructions is negligible. 
The dashed line in the upper panel represents the average of
$\delta$ values in the upper-limit case at separations larger than 20
Mpc, no longer affected by the ionising flux from the QSO. The dotted
lines are the $1$ and $2\,\sigma$ standard deviations and the shaded
region represents the $3\,\sigma$ level. 
A significant decrease of the density field is observed in the first 
bin at proper separation from the emitting QSO of less than 8 Mpc,
corresponding to the average radius of influence of the QSOs in our
sample. 



\subsection{The density structure around QSOs}
\label{sect_PEAK}

\begin{figure*}
\begin{center}
\includegraphics[width=16cm]{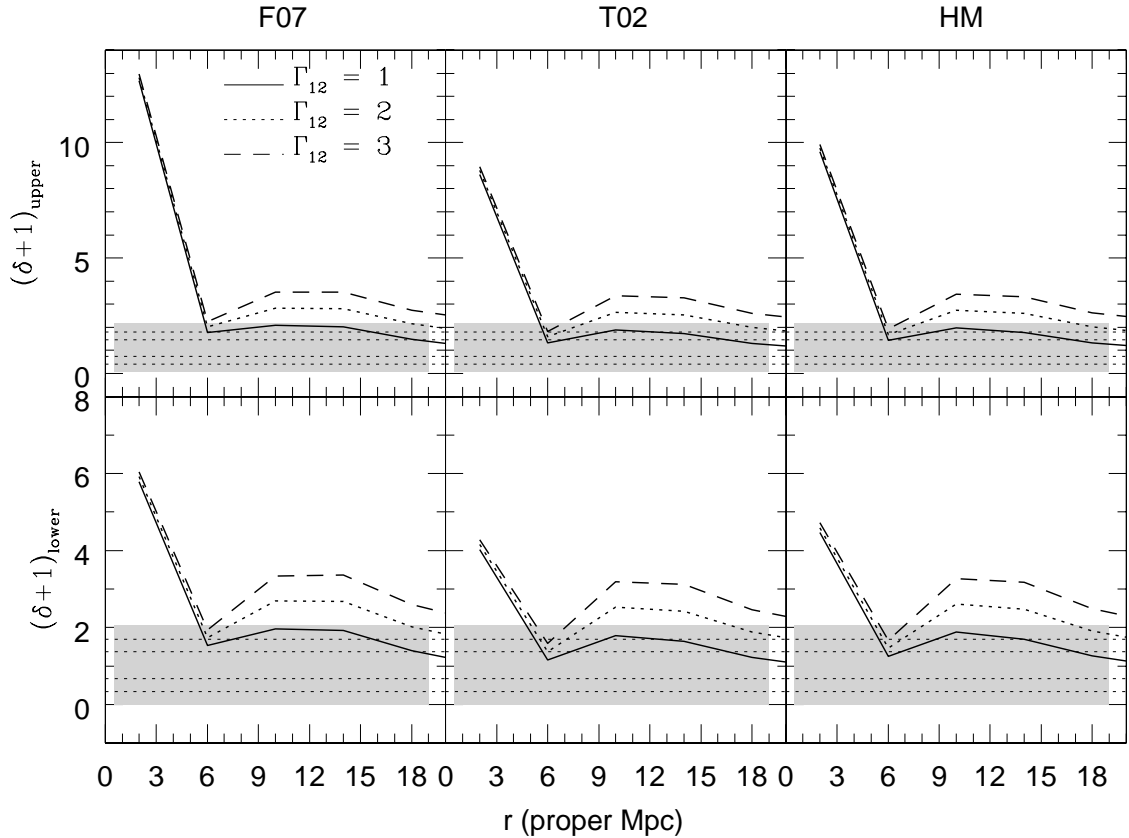}
\caption{{\protect\footnotesize{Mean $\delta$ field in the upper (top
      panels) and lower (bottom panels) limit case in the region close
      to the emitting QSO (which is located 
      at $r=0.0$) for the QSO spectra in our sample. In each row,
      the three panels differ for the adopted QSO continuum slope (see
      Section~\ref{sub_lum}). The three curves in each panel differ
      for the adopted value of $\Gamma_{12}$. 
      The $3\,\sigma$ fluctuations in the IGM region (far away from
      the QSO) are represented by the shaded area.}}} 
\label{grid_QSO}
\end{center}
\end{figure*}

In order to recover the density field in the region close to the
emitting QSO, its ionising flux has to be added to the UV background, 
and the total ionisation rate has to be used in eq.~\ref{tra} of the
{\small FLO} algorithm. We define $\Gamma_{12}'$ as: 

\be
\Gamma_{12}'= \frac{\Gamma_{\rm UVB}+\Gamma_{\rm QSO}}{10^{-12}} 
\ee
where $\Gamma_{\rm QSO}$ was defined in eq.\ref{eqGQ}. As explained in
Section~\ref{sub_lum}, the value of $\Gamma_{\rm QSO}$ depends on the
adopted slope for the QSO continuum in the region blue-ward of 
the Lyman limit. We investigated how the three different assumptions
for the slope influenced the final result.  

In Fig.~\ref{grid_QSO}, we present the mean $\delta$ field for the
lower and the upper limit case recovered from the spectra in our
sample in the region within 20 Mpc from the radiating source in bins of 4 proper Mpc. 
We studied how the {\small FLO} output was modified by assuming different
ionisation rates for the UV background, $\Gamma_{12}=1,\ 2,\ 3$, and
different slopes for the QSO continuum. 

The first important result of our analysis is that {\small FLO}
recovers the over-density hosting the QSOs in each one of the panels
of Fig.~\ref{grid_QSO}.
The second result is that the value of $\delta$ for the peak varies
significantly depending on several factors that are discussed in
the following.  

\begin{description}

\item[1.{\em Treatment of the empty bins:}] at variance with the region away
from the QSO, the lower and upper limit case in the treatment of the
empty bins makes a difference of a factor of $\sim 2$ for the $\delta$
value of the peak of the over-density where the QSO resides. This is
due to the QSO strong ionising field that, when applied in
eq.~\ref{tra}, transforms even small column densities into large
over-densities. Increasing the SNR of our spectra would decrease the
minimum detectable \HI\ column density, thus decreasing the
difference between the two limiting cases. 

\item[2. {\em QSO continuum slopes:}] the difference between the two
  extrapolations with the standard fixed power laws, with indexes
  $\gamma=1.5$ and 1.8, is $\sim 10$ percent, while there is
   an increase in the peak value between $\sim 30$ and 45
   percent when the F07 slopes are adopted. In the latter model, the
   continuum in the blue is obtained by extrapolating the slope
   red-ward of the \Lya emission, so it can be considered as a sort of
   upper-limit to the true continuum, that cannot be recovered due to
   contamination by the IGM absorption. 

\item[3. {\em Ionisation rate of the UVB:}] this parameter
does not have any influence on the $\delta$ value for the QSO peak, as
expected. On the other hand, the effect on the average value of the
$\delta$-field away from the QSO is significant and will be discussed
in Section~\ref{sub_gamma}.
    
\item[4. {\em QSO systemic redshifts:}] the uncertainties in the
  emission redshifts of the QSOs whose spectrum was used to determine
  the over-density are of the order of $100-200\ \kms$ corresponding
  to $\sim 400-800$ proper kpc at $z \simeq 2.5$. As a consequence, we
  do not expect a major influence on the value of the QSO over-density
  which is computed in a bin of 4 proper Mpc. A significant
  improvement in the result would be obtained if the systemic redshift of
  all the QSOs in the sample would be measured with intermediate
  resolution spectra in the infrared. 

\begin{figure}
\begin{center}
\includegraphics[width=8cm]{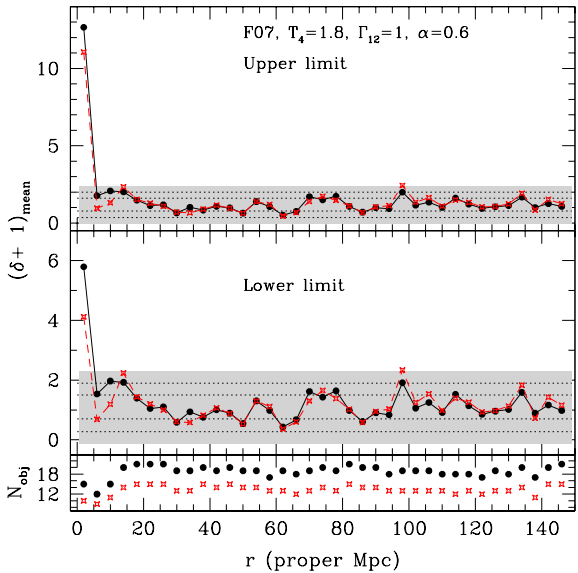}
\caption{{\protect\footnotesize{Comparison between the recovered
      average density field for the total sample (solid dots) and the
      sample without the QSOs with associated systems (empty
      stars) in the reference model with the F07 slopes. The shaded
      regions mark the $3\,\sigma$ interval an the dotted lines the 1
      and $2\,\sigma$ level for the reduced sample. In the lower
      panel the number of objects for the two 
      samples is reported. }}} 
\label{noaal}
\end{center}
\end{figure}

\item[5. {\em Associated absorption systems:}] metal absorption
  systems within $\sim 5000$ km s$^{-1}$ (corresponding to $\sim 20$
  proper Mpc) from the emission redshift of the QSO are defined as
  {\sl associated} to the QSO. Some of these systems could indeed be
  intrinsic to the QSO, that is they could be very close to the
  emitting region and their position could be determined only by the
  large velocity at which they were ejected from the QSO
  itself. Counting these lines as intervening would result in an
  over-estimate of the density field in the proximity of the QSO. Six
  QSOs in our sample show associated absorption lines
  (AAL). Figure~\ref{noaal} reports the comparison between 
the recovered density field of the total QSO sample and of the sample
without AAL for the reference model and the F07 slopes. There is a
decrease of $\sim 40$ and $15$ percent in the $\delta$ value of the
peak in the lower and upper limit-case, respectively. A slight
increase in the scatter of the IGM density field is also measured due
to the lower statistics. The point at a separation of 98 Mpc is above
the $3\,\sigma$ level, however the fluctuation is due to a single QSO,
PKS1448-232 and disappears when the QSO is eliminated from the
sample.
\end{description}

\subsection{Comparison with previous results}
\label{sub_comp}

The density distribution in the region close to bright QSOs was
studied in two previous works \citep{rollinde,guimaraes} with a
different technique: the variation of the cumulative probability
distribution function (CPDF) of the optical depth measured in the \Lya
forest region away from and close to the QSO.
The advantage of this method with respect to {\small FLO} is that the
fitting of the lines is not necessary since the observed quantity used
is the pixel-by-pixel transmitted flux.      
On the other hand, before investigating the change of the CPDF with the
distance from the QSO, it is necessary to evaluate its evolution with
redshift and subtract it. Then, the change in the CPDF has to be
translated into a variation of the density field with a bootstrap
technique \citep[see][for the detailed description]{rollinde}.   

\citet{rollinde}, using a subsample of our sample formed by 12 QSO
spectra with $\langle z_{\rm em} \rangle \simeq 2.5$, found a
significant over-density for separations 
between $\sim3$ and 15 proper Mpc assuming $\Gamma_{12}<3$. They
assumed in the computation a QSO continuum with a fixed power law
slope $\gamma=0.5$ and a temperature-density relation index
$\alpha=0.5$. 
\citet{guimaraes} applied the same technique to a sample of 45 QSO
spectra at intermediate resolution and at redshift $\langle z_{\rm em}
\rangle\sim 3.8$. An over-density extending to separations of $\sim 15$
proper Mpc is detected adopting the parameters, $\Gamma_{12}=1$, $T_{0,4}=1$,
$\alpha=0.0$ and $\gamma=0.7$. These authors claimed also the
detection of a correlation between the over-density and the luminosity:
brighter QSOs reside in higher over-densities. 

The main difference between our result and the previous ones is that we
clearly detected an over-density limited to a few Mpc from the
QSO, while the results of the optical depth analysis showed a smooth
decrease from the peak to the IGM density extending for more than 10
Mpc. 
In particular, the difference with the result at higher redshift by
\citet{guimaraes} cannot be ascribed to a difference in QSO intrinsic
luminosity, since the average luminosity of the two samples is the
same: $\log L_{\rm LL} \simeq 31.7$.
We will see in Section~\ref{sect_SIM} that the cosmological
hydro-simulations that we used for comparison show very narrow density
peaks as found in our recovered density field.  
Very similar results were obtained by \citet{fg} who investigated with
hydrodynamical simulations the bias introduced by the QSO over-density
in the estimate of the UV background intensity from the proximity
effect. They considered three ranges of masses
$3.0\pm 1.6 \times 10^{11}$ M$_{\odot}\ h^{-1}$,  $3.0\pm1.6\times 10^{11}$  
M$_{\odot}\ h^{-1}$ and $6.0\pm 3.2\times 10^{12}$
M$_{\odot}\ h^{-1}$ for the DM haloes hosting QSOs in the simulations
and averaging the results of 100 lines of sight obtained overdensity
profiles extending to $\sim 3-5$ proper Mpc at redshifts $z\simeq2-3$.

\subsection{Constraints on the IGM physical parameters}
\label{sub_gamma}

The {\small FLO} algorithm depends on  the physical parameters
of the IGM, in particular the temperature at the average density, the
UVB ionisation rate and the index of the temperature-density
relation. We adopted for these parameters the values from the
cosmological hydro-simulation used to validate the {\small FLO}
performances in paper I. In turn, these values are in agreement with
the most recent observational measurements
\citep[e.g.][]{ricotti,schaye00,scott00,tytler04,bolton05}.  

In Fig.~\ref{grid_QSO}, we showed that values of the UVB ionisation
rate $\Gamma_{12}\ge 2$ (corresponding to UVB intensities at the
Lyman limit, $\log J_{\rm LL} \ge -21.15$) are rejected at more than
$3\,\sigma$ because they over-estimate the average density of the
IGM. 

\begin{figure*}
\begin{center}
\includegraphics[width=16cm]{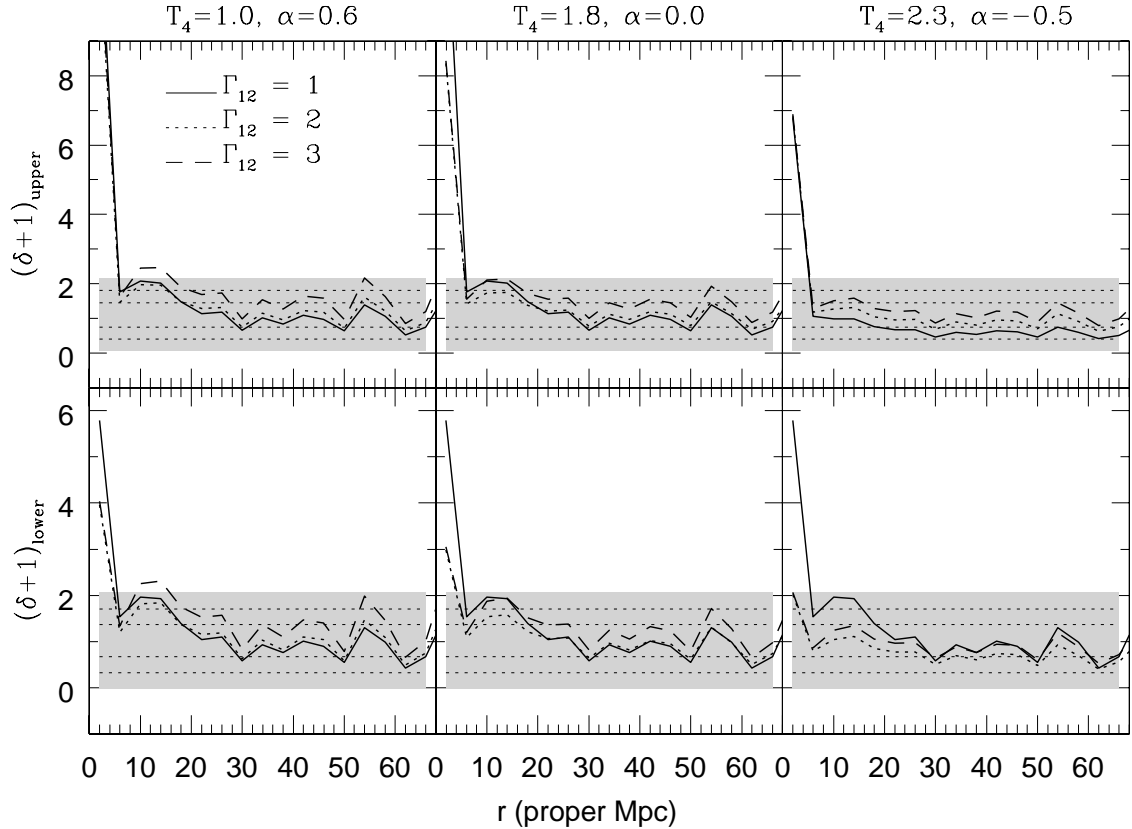}
\caption{{\protect\footnotesize{Variation of the {\small FLO} output
      $\delta$-field with the physical parameters of the IGM. The
      solid line represents the fiducial model with $T_{0,4}=1.8$,
      $\Gamma_{12}=1$ and $\alpha=0.6$ with the F07 QSO slopes. The
      dotted and dashed lines are obtained for $\Gamma_{12}=2$ and
      $\Gamma_{12}=3$, respectively, and varying the
      temperature and/or the index $\alpha$ as marked above the
      panels. The shaded regions and the dotted horizontal lines are
      the same of Fig.~\ref{grid_QSO}.}}} 
\label{grid_QSObis}
\end{center}
\end{figure*}

We investigated how the variation of the other two relevant
parameters, $T_{0,4}$ and $\alpha$, influence the reconstruction of
the $\delta$-field with {\small FLO}. The results are shown in
Fig.~\ref{grid_QSObis}. A temperature at the average density almost a
factor of two lower than the reference one, which implied a decrease
in the smoothing scale of $\sim 30$ percent, could reconcile values of
$\Gamma_{12}=2$ with the IGM average density but not values as large
as $\Gamma_{12}=3$. However, this temperature is at the lowest end of
the measured temperatures in the IGM at the considered redshifts. 
Higher temperatures and high values of $\Gamma_{12}$ allow to recover
the correct IGM average density if a lower $\alpha$ index is adopted. 
In particular, an inverted temperature-density relation with
$\alpha\sim-0.5$, as measured by \citet{bolton08} from the flux
probability distribution function, would require a high
temperature, and a high UVB ionisation rate to recover the average IGM
density.   

However, the validity of the {\small FLO} reconstruction with a different set
of parameters than the reference one needs to be verified with
numerical simulations and we plan to do it in the next paper of the
series.

\section{Comparison with simulations}
\label{sect_SIM}

We compared the reconstruction of the QSO over-density by {\small FLO}
with the average density field from a cosmological hydro-simulation in
the proximity of the halos that would likely harbour the QSOs in our
sample.  
 
\subsection{Simulated lines of sight}

The simulations were run with the parallel hydro-dynamical (TreeSPH)
code {\small {GADGET-2}} based on the conservative
`entropy-formulation' of SPH \citep{gadget}. They consist of a
cosmological volume with periodic-boundary conditions filled with an
equal number of dark matter and gas particles.  Radiative cooling and
heating processes were followed for a primordial mix of hydrogen and
helium.  We assumed a mean UVB produced by QSOs and
galaxies as given by \citet{HM} with helium heating rates multiplied
by a factor 3.3 in order to better fit observational constraints on
the temperature evolution of the IGM \citep[][]{schaye00}. This
background gives naturally 
a $\Gamma_{\rm UVB}\sim 10^{-12}$ at the redshifts of interest here \citep{bolton05}.  
The star formation criterion is a very simple one
that converts in collision-less stars all the gas particles whose
temperature falls below $10^5$ K and whose over-density is larger than
1000. 
More details can be found in \citet{viel04}.  
The cosmological model corresponds to a `fiducial' $\Lambda$CDM
Universe 
\citep[the B2 series of][]{viel04}.

We used $2\times 400^3$ dark matter and gas particles in a $120\
h^{-1}$ comoving Mpc box.
Having a large box size is crucial since the influence zone of the
QSOs is usually of the order of some Mpc or tens of Mpc (see
Table~\ref{tab1} of this study). 
The gravitational softening was set to 5 $h^{-1}$ kpc in comoving units for 
all particles.
We analysed the output at $z=2.2$ and run a Friend-of-Friend (FoF)
algorithm to identify the most massive collapsed haloes that should
host the QSOs.  We found about 400 (54) haloes whose total mass is
larger than $2\times 10^{12}$ M$_{\odot}\ h^{-1}$ (10$^{13}$
M$_{\odot}\ h^{-1}$). We then pierced the simulated box along the 400 lines
of sight (LOSs) intersecting the centre of the haloes with M~$\ge 2
\times 10^{12}$ M$_{\odot}\ h^{-1}$ and along random directions.  This
latter sample constitutes our `control' sample.  We explicitly checked
that the correlation function of the haloes with masses larger than
$2\times 10^{12}$ M$_{\odot}\ h^{-1}$ is reasonably well fitted by a
power-law function with $r_0=6 \mpc\ h^{-1}$ and slope $-1.8$ in agreement
with observational results \citep{croom05}. 
Furthermore, it was recently found that QSOs could be typically hosted in
haloes of mass $3\times 10^{12}$ M$_{\odot}\ h^{-1}$ regardless of
their luminosity and redshift \citep{daangela}.   


\begin{figure}
\begin{center}
\includegraphics[width=8cm]{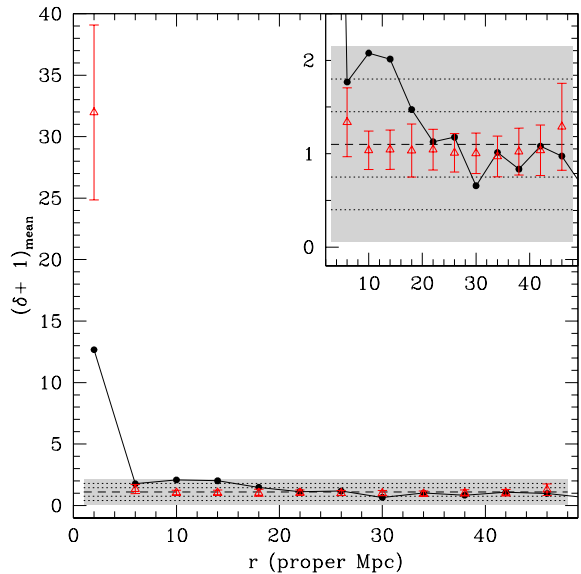}
\caption{{\protect\footnotesize{Comparison between the {\small FLO}
      average density field for the observed QSO sample (solid dots, upper
      limit case) in the reference model with the F07 slopes and the
      density field from simulations (empty triangles with $1\,\sigma$
      error bars). The window in the upper right corner is a zoom of
      the fluctuations at the mean level. The shaded area marks the
      $3\,\sigma$ regions, the dotted lines mark the 1 and
      $2\,\sigma$ and the dashed one the mean density level.}}} 
\label{sim}
\end{center}
\end{figure}

\subsection{Comparison of the density fields}

For each one of the 400 simulated lines of sight, the density
contrast, the temperature, and the peculiar velocity are known pixel
by pixel. Peculiar velocities are small, typically less than 100 $\kms$,
and randomly oriented. However, since the observed density fields were
recovered in redshift space, the mock lines of sight were 
modified by correcting the redshifts of the density
field ($z_{\rm old}$) with the peculiar velocity field to obtain the
density field in redshift space ($z_{\rm new}$) using the formula
$v_{\rm pec}(z_{\rm old}) = c\, (z_{\rm new} - z_{\rm old})/ (1 +
(z_{\rm new} + z_{\rm old})/2)$ and the periodic-boundary conditions.
Then, the mock LOSs were shifted (applying the periodic-boundary
condition) in order to have the peak of the
most massive halo (where the QSO should reside) in the first pixel and
the proper separation from this pixel was computed for all the other
pixels. The total length of the simulated LOSs is $\sim 50$ proper
Mpc. The obtained density fields were finally rebinned into steps of 4
proper Mpc as in the case of the observed ones.    
The final simulated density field was computed by averaging in each
bin the mean $\delta$ of 100 samples of 19 LOSs (the average number
of objects per bin in the observed sample) extracted from the total
sample of 400 mock LOSs with a bootstrap technique. 
The error bars on the final density field
were computed from the standard deviation of the 100 samples.    

The comparison between the simulated density field and the one recovered
from observed spectra is plotted in Fig.~\ref{sim}. The simulated
$\delta$ field is consistent with the mean IGM density down to 4 Mpc
and up to 48 Mpc (see zoomed window). 
The disagreement between the simulated and observed density contrast
in the first bin and, in particular, the large value of $\delta$
obtained from the mock spectra, could be explained by the fact that
the considered simulations do not include in the most massive halos
the presence of the AGN that would accrete part of the gas in the peak.

\section{Conclusions}
\label{sectCONC}

In this work, we used a sample of $\sim 6300$ \Lya\ absorption lines
obtained from high resolution, high signal-to-noise ratio spectra of
21 QSOs in 
the redshift range $2.1 \la z_{\rm em} \la 3.3$ and  the {\small FLO}
algorithm for the reconstruction of baryon density fields (described
in details in Paper I) to investigate the baryon distribution as a
function of the distance from the QSO. 

The effect of the QSO radiation (supposed to be isotropic) dominates
over the UV background (UVB) in a sphere with radius varying from
$\sim 3$ to 15 Mpc for our sample, with an average of $\sim 8$ Mpc. 
The increased ionisation flux modifies the local line number density
causing the so-called {\sl proximity effect}, that was used in the past to
estimate the intensity of the UVB at the Lyman limit. However, the
local line number density is determined not only by the ratio between
the UV flux from the QSO and the UVB but also by a variation of the
density field due to the fact that QSOs likely lie on density
peaks. 

The recovery is extremely sensitive to the emission redshift and to
the UV flux of the QSOs. 
We spent a significant effort to obtain, both from
the literature and from the spectra at our disposal, the best estimates 
for the systemic redshifts of the QSOs in our sample in order to
reduce the uncertainties to less than $100-200\ \kms$. For the 16 QSOs
for which the redshift was determined with $H_{\beta}$, [\OIII] (5
QSOs) or with the \OI, \MgII\ UV emission lines, the average 
velocity difference with respect to the determinations present in the
literature \citep[e.g.][ based mainly on the high-ionisation UV lines
  and \Lya]{kim04} is $\sim -500\ \kms$. 
For the QSO UV flux, we adopted three different models: two with a
power law with a fixed slope and one for which each QSO has its own
slope obtained from the comparison with a spectral library. The
resulting ionisation rates for the three models are within a factor
of two.   

We improved the performances of {\small FLO} with respect of paper I
by adopting a smoothing scale $\sim 30$ percent smaller than the Jeans
length, following the prescription by \citet{gnedin_hui}. This
contrivance allows us to solve the problem of over-estimation of the
over-densities we had in paper I and to correctly recover the mean
density in the IGM.   

We applied the {\small FLO} algorithm to each QSO spectrum  
and recovered the sample-averaged $\delta$ field as a function of the
distance from the emitting sources, in bins of 4 proper Mpc.  
The obtained results are described in the following sections. 

\subsection{QSO over-density from the proximity effect}

\begin{description}
\item[i)] If only the UVB ionisation rate is adopted in eq.\ref{tra}
  of {\small FLO}, neglecting the QSO radiation, a decrease of the 
  mean density field significant at $\sim 2\,\sigma$ is observed
  within 8 proper Mpc of the emitting QSO, confirming the presence of
  a proximity effect. 

\item[ii)] When also the QSO ionisation rate is taken into account an
  over-density significant at more than $3\,\sigma$ is recovered at
  proper separations of less than 4 Mpc (first bin).

\item[iii)] The absolute $\delta$-value of the QSO over-density is
  uncertain by an overall factor of $\sim 3$. A factor of $\sim 2$
  variation is due to the lower and upper-limit 
  $\delta$-value derived for the empty bins during the field
  reconstructions: corresponding to no absorber or to one absorber
  with the minimum \HI\ column density detectable in our sample, respectively. 
  This uncertainty can be improved by using more spectra with a larger
  signal-to-noise ratio. Another factor of $\sim 1.1-1.5$ is due to the
  different assumptions for the QSO continuum in the region blue-ward
  of the \Lya\ emission.

\item[iv)] The uncertainty on the systemic redshift of the QSOs used 
  to determine the density distribution in their neighbourhood is $\la
  200\ \kms$ corresponding to separations $\la 800$ proper kpc. As a
  consequence, it should have negligible effects on the $\delta$-value  
  of the peak. 

\item[v)] We compared the resulting density field with the average density
  field in the proximity of massive haloes ($M>2\times 10^{12}$
  M$_{\odot}\ h^{-1}$) in a cosmological hydro-simulations. The two
  distributions are in reasonably good agreement.

\item[vi)] There is a significant discrepancy between our results and
  the previous determinations of the matter distribution around QSOs, 
  obtained with the optical depth statistics \citep[ODS,
  ][]{rollinde,guimaraes} at redshifts $\langle z_{\rm em}\rangle
  \simeq 2.5$ and 3.8. The ODS method recovers gaseous
  over-densities extending to scales as large as $\sim 15$ Mpc while our
  overdensity is limited to a region closer than 4 proper Mpc from the
  QSO. We would need to increase our sample, in particular with QSOs
  without associated systems in order to study if the brightest
  objects resides in more extended peaks.
\end{description}

\subsection{Constraints on the IGM physical parameters}

\begin{description}
\item[i)] In the hypothesis of a temperature of the gas at the mean
  density $T_0=1.8\times 10^4$ K and an index of the temperature-density
  relation for the IGM $\alpha=0.6$, an UVB ionisation rate of 
  $\Gamma_{\rm UVB} \simeq 10^{-12}$  s$^{-1}$ gives the correct IGM 
  mean density. On the other hand, $\Gamma_{\rm UVB} \simeq 2\times 
  10^{-12}$ and $3\times 10^{-12}$ s$^{-1}$ are
  excluded at more than 2 and $3\,\sigma$, respectively, because the
  recovered {\small FLO} IGM density field overestimates the mean
  density. 

\item[ii)] Values of $\Gamma_{\rm UVB} > 10^{-12}$ s$^{-1}$ can be
  reconciled with the correct IGM mean density if different
  combination of $T_0$ and $\alpha$ are adopted.  In particular, an
  inverted temperature-density relation with $\alpha\simeq -0.5$ used
  in the {\small FLO} algorithm gives the correct IGM mean density  
  for $\Gamma_{\rm UVB} \simeq 3\times 10^{-12}$ s$^{-1}$ and
  $T_0\simeq 2.3\times10^4$ K. Such large values of the ionisation
  rate could arise as a consequence of the \HeII\ reionisation at
  $z\sim3$ and of UVB fluctations \citep[e.g.][]{meiksin_white}. 
  The performances of {\small FLO} with 
  different set of parameters than the reference one have however to
  be tested with numerical simulation. We defer the details of this
  analysis to a future paper.   

\end{description}
    
\subsection{Limiting factors and future developments}

\begin{description}
\item[i)] Most of the high redshift QSOs have redshifts determined
  from UV emission lines which are known to be systematically shifted
  with respect to systemic redshifts. This is particularly critical
  for proximity effect studies both along and transverse to the line
  of sight. An improvement in this sense is expected from the new
  intermediate-resolution, UV to near-IR spectrograph X-Shooter at the
  VLT \citep{vernet}, that will be operative from the first trimester of
  2009.
\item[ii)] The temperature of the IGM gas at the mean density is
  highly uncertain. Its best determinations date back to 2000, the
  evidence of a jump in its value at $z\sim 3$ was marginal and should 
  be verified with the present larger samples of
  high-resolution, high-signal-to-noise QSO spectra. Furthermore,
  those temperature estimates are used in many cosmological
  hydro-simulations to re-normalise the ionising background intensity.
\item[iii)] The intensity and nature of the UV background is another unsolved
  riddle which should require a new observationally-based
  determination (with the caveat in i) since the present simulations
  are only now starting to have the resolution and the physics
  (e.g. the radiative transfer) needed to derive it
  self-consistently. 
\end{description}

\section*{Acknowledgments}   
We are grateful to A. Grazian for useful discussions and to an
anonymous referee for his/her constructive comments. 
It is a pleasure
to thank P. Marziani and collaborators for having shared the
information on the systemic redshift of 3 QSOs of our sample before
publication.  Numerical computations were done on the COSMOS
supercomputer at DAMTP and at High Performance Computer Cluster (HPCF)
in Cambridge (UK). COSMOS is a UK-CCC facility which is supported by
HEFCE, PPARC and Silicon Graphics/Cray Research.

\label{lastpage}
\newpage
\end{document}